\documentclass[twocolumn,prc,aps,showpacs,superscriptaddress,amsmath,amssymb,floatfix,nofootinbib,superscriptaddress]{revtex4}
\usepackage[usenames]{color}
\usepackage{graphicx}
\usepackage{dcolumn}
\usepackage{bm}

\begin{document}


\title{Pion-induced Drell-Yan processes and the flavor-dependent EMC effect}

\author{D.~Dutta}
\affiliation{Department of Physics, Mississippi State University, 
Mississippi State, Mississippi 39762, USA}
\author{J. C.~Peng}
\affiliation{Department of Physics, University of Illinois at 
Urbana-Champaign, Urbana, Illinois 61801, USA}
\author{I. C. Clo\"{e}t}
\affiliation{Department of Physics, University of Washington, Seattle, 
Washington 98195, USA}
\author{D.~Gaskell}
\affiliation{Thomas Jefferson National Accelerator Facility, Newport News, 
Virginia 23608, USA}


\begin{abstract}
Pion-induced Drell-Yan processes are proposed as a potential tool to measure the 
flavor dependence of the EMC effect, that is, the flavor-dependent modification 
of quark distributions in the nuclear medium. 
Existing pionic Drell-Yan data are compared with calculations using 
a recent model for nuclear quark distributions that incorporates
flavor-dependent nuclear effects. While no firm conclusions can yet be drawn, we 
demonstrate that existing Drell-Yan data seem to imply a flavor dependence of 
the EMC effect. We highlight how pion-induced Drell-Yan experiments on nuclear targets
can access important new aspects of the EMC effect, not probed in deep inelastic scattering,
and will therefore provide very stringent constrains for models of nuclear quark distributions. 
Predictions for possible future pion-induced Drell-Yan experiments are also presented. 
\end{abstract}


\maketitle

The depletion of the nuclear structure functions in the valence quark region
was discovered in 1983 by the European Muon Collaboration (EMC) 
in a muon-induced deep inelastic scattering (DIS) experiment~\cite{emc1},
and is now known as the EMC effect.
The EMC effect
provides clear evidence that the quark distributions in nuclei are modified 
compared to those of free nucleons. The observation of the EMC effect was 
unexpected, since the nuclear binding energy is orders of magnitude smaller than 
the energy scale probed in DIS. Other DIS experiments using electron~\cite{slac}, 
muon~\cite{emc1,muons} and neutrino~\cite{neutrinos} beams have now confirmed the 
EMC effect over a broad range of nuclear masses ($A$) and momentum transfers ($Q^2$). 
The improved precision of these experiments, including the recent experiments on light 
nuclei~\cite{hermes,seely09}, have provided a detailed and multi-dimensional view of 
the nuclear modification of the quark distributions. The EMC effect has also 
been experimentally verified in the time-like region using both the pion- and 
proton-induced Drell-Yan processes~\cite{na10,drell_yan,dyreview}.

Despite a quarter century of significant experimental and theoretical effort, the 
specific origins of the observed $A$ dependence of the nuclear quark distributions
have yet to be unambiguously identified. Attempts to explain the 
EMC effect have led to a large collection of theoretical models~\cite{geesaman_review,norton_review}, 
many of which are
capable of describing the essential features of the data, however the underlying 
physics mechanisms in each model are often very different.
Therefore, it appears likely that before the true origins of the EMC effect 
are understood new experiments are necessary, which explore aspects of the EMC 
effect not effectively probed in DIS, and will therefore 
help distinguish between the various models.
In this Letter we explore the effectiveness of pion-induced Drell-Yan processes 
as a tool to investigate new aspects of the EMC effect~\cite{Ericson:1984vt,Bickerstaff:1985ax}, 
in particular its flavor dependence.

A new calculation of the modifications of nucleon quark
distributions in the nuclear medium has recently been reported~\cite{ianemc,ian}. 
In this approach, by Clo\"{e}t, Bentz and Thomas (CBT), the Nambu--Jona-Lasinio
model is used to describe the coupling of the quarks in the bound nucleons 
to the scalar and vector mean fields inside a nucleus. 
These nucleon quark distributions are then
convoluted with a nucleon momentum distribution in the nucleus to generate the
nuclear quark distributions~\cite{ianemc}. An important feature of this model
is that for $N\neq Z$ nuclei (where $N$ and $Z$ refer to the number of
neutrons and protons) the isovector--vector mean field (usually denoted by $\rho^0$) 
will affect the up quarks
differently from the down quarks in the bound nucleons. Therefore, this model
has a novel prediction that the $u$ and $d$ quarks have distinct nuclear
modifications for $N\neq Z$ nuclei. In other words, the EMC effect is
flavor sensitive and depends on the $N/Z$ ratio of the nucleus~\cite{ianemc}.
This flavor-dependent EMC effect was also shown~\cite{ian,Bentz:2009yy} to account
for a large fraction of the discrepancy between the NuTeV collaboration
measurement~\cite{nutev} of the weak mixing angle and the world average.

\begin{figure}[tbp]
\centering\includegraphics[width=\columnwidth]{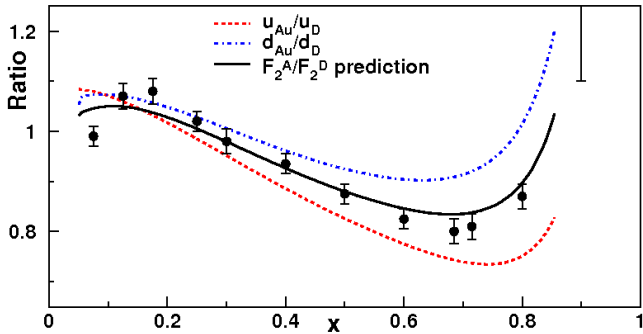}
\caption{Ratios of quark distributions and structure functions in
nuclear matter 
versus the deuteron plotted as a function of Bjorken-$x$, at $Q^2 = 10\,$GeV$^2$. 
The solid circles are data for $N=Z$ nuclear matter from Ref.~\cite{day-sick}. The
solid curve is the calculation of $F_2^A/F_2^D$ for $N=Z$ nuclear matter from
Clo\"{e}t, Bentz, and Thomas~\cite{ianemc,ian}. The dashed and dot-dashed 
curves are the ratios of quark distributions in a gold nucleus to 
those in a deuteron, for $u$ and $d$ quarks, respectively.}
\label{fig1}
\end{figure}

Figure~\ref{fig1} shows the prediction for the nuclear dependence 
of quark distributions in the CBT model. 
The solid curve is the result for the usual EMC effect in symmetric ($N=Z$)
nuclear matter, that is, the ratio $F_2^A/F_2^D$, where $F^A_2$ and $F_2^D$
are the per-nucleon structure functions of the nucleus and the deuteron,
respectively.\footnote{The CBT results used throughout this paper are calculations 
performed for asymmetric nuclear matter. Results for particular nuclei are obtained
by choosing the $N/Z$ ratio equal to the nucleus in question. The
deuteron results are obtained from a combination of free proton and neutron quark
distributions.}
The calculation is in good agreement with the data extracted by Sick and 
Day~\cite{day-sick}. For an isoscalar nucleus the $u$ and $d$ quark distributions
are modified in the same manner, however
for a $N \neq Z$ nucleus the $u$ and $d$ quark distributions are 
predicted to undergo different modifications in the CBT model. 
For example, the dashed and dot-dashed
curves in Fig.~\ref{fig1} are, respectively, the predicted ratio of the $u$- and 
$d$-quark distributions in a gold nucleus to those in the 
deuteron.
For an $N > Z$ nucleus (such as gold), the $\rho^0$
mean field leads to a stronger nuclear binding for the $u$ quarks compared to that 
for the $d$ quarks~\cite{ianemc}. Hence, the medium modification of the $u$-quark 
distribution is enhanced, while the $d$-quark distribution is modified less by the
medium, as illustrated in Fig.~\ref{fig1}.

The flavor dependence of the EMC effect is a
promising experimental observable to distinguish
among the plethora of models that can describe the EMC effect. In the
simplest picture, we can consider the nuclear parton distribution
functions (PDFs) to be just the nucleon PDFs smeared by the Fermi motion
of the nucleons. In this picture it is natural that the nuclear
modification would be very similar for $u$ and $d$ quarks.
Similarly, the $Q^2$-rescaling model~\cite{rescaling} also predicts 
no quark-flavor dependence
for nuclear modifications. In contrast, the pion-excess 
model~\cite{pion1,pion2,pion3} would naturally
predict flavor-dependent nuclear modifications arising from the different 
isospin composition of the pion cloud components of protons and neutrons.

Since the inclusive DIS experiments measure the combined nuclear
modifications of $u$ and $d$ quarks, it is difficult to extract 
the quark-flavor dependence of the EMC effect in these experiments. 
However, it has been demonstrated~\cite{ian_new} that the 
parity violating DIS asymmetry, $A_{PV}$, is very sensitive to the 
quark-flavor dependence of the EMC effect.  Measuring $A_{PV}$ for 
a variety of $N \neq Z$ nuclei has the potential to provide a clean demonstration 
of flavor-dependent nuclear modification of the the quark distributions. 
Such a measurement is part of the large parity violating DIS program 
proposed~\cite{pr12007} at the upgraded $12\,$GeV JLab facility currently under construction. 
   
Semi-inclusive DIS (SIDIS) on heavy nuclear targets, 
in which the flavor of the struck quark is tagged by the detected hadron, is also a 
promising experimental tool to search for the flavor-dependent EMC effect. 
Recently, Lu and Ma~\cite{ma2006} pointed out that charged lepton SIDIS off nuclear 
targets and the deuteron can be used to probe the flavor content of the nuclear 
quark sea, which can help distinguish between the various models of the EMC effect. 
Indeed, a SIDIS experiment~\cite{pr12004} aiming at a precise 
determination of flavor dependence of the EMC effect has also been proposed at the 
upgraded $12\,$GeV JLab facility. 

The focus of this Letter is pion-induced Drell-Yan processes, which are 
complementary to the DIS processes mentioned above and provide another 
experimental tool with which search for flavor-dependent effects in the 
nuclear modification of the nucleon structure functions. 
First, we will identify the experimental observables sensitive to the
flavor-dependent EMC effect. Then we compare existing pion-induced
Drell-Yan data with calculations using the CBT model, and
examine how well the data can constrain the flavor dependence of the 
EMC effect. Finally, predictions for possible future pion-induced 
Drell-Yan experiments are presented.

The leading order Drell-Yan cross-section for a pion interacting with
a nucleus is given by
\begin{multline}
\frac{d\sigma^{2}_{\pi^{\pm} A}}{dx_{\pi}\,dx_2} = \frac{4\pi\alpha^2} {9\, s\,x_{\pi}\,x_2} \\
\times \sum_q e_q^2
\left[q_{\pi^{\pm}}\!(x_{\pi})\,\bar{q}_A(x_2) + \bar{q}_{\pi^{\pm}}\!(x_{\pi})\,q_A(x_2)\right],
\label{eq:crosssection}
\end{multline}
where $\alpha$ is the fine structure constant, $s$ is the center of mass energy 
squared, $x_{\pi}$ is the Bjorken scaling variable for the interacting quark in the pion
and $x_2$ is the analogous quantity for the nucleon in the target nucleus.
Quark flavor is labeled by $q$, where $e_q$ is the quark charge, the 
pion quark or antiquark distributions are labeled by $q_{\pi^{\pm}}$ 
and $\bar{q}_{\pi^{\pm}}$, respectively, and the subscript
$A$ indicates nuclear quark or antiquark distributions.

To explore the sensitivity of pion-induced Drell-Yan processes to 
a flavor-dependent EMC effect, we consider the three ratios 
$\frac{\sigma^{DY}(\pi^++A)}{\sigma^{DY}(\pi^-+A)}$, $\frac{\sigma^{DY}(\pi^-+A)}
{\sigma^{DY}(\pi^-+\text{D})}$ and 
$\frac{\sigma^{DY}(\pi^-+A)}{\sigma^{DY}(\pi^-+\text{H})}$,
where $A$ represents a nuclear, D a deuteron and H a hydrogen target.
Assuming isospin symmetry, which implies
$u_{\pi^+} = d_{\pi^-}$, $\bar{u}_{\pi^-} = \bar{d}_{\pi^+}$,
$\bar{u}_{\pi^+} = \bar{d}_{\pi^-}$, $u_{\pi^-} = d_{\pi^+}$ and
keeping only the dominant terms in each cross-section, one readily obtains
%
\begin{align}
\label{eq:Rpm}
R_{\pm} &= \frac{\sigma^{DY}(\pi^+ + A)}{\sigma^{DY}(\pi^- + A)}
\approx \frac{d_{A}(x)}{4\,u_{A}(x)}, \allowdisplaybreaks \\
R^{-}_{A/D} &= \frac{\sigma^{DY}(\pi^- + A)}{\sigma^{DY}(\pi^- + \text{D})}
\approx \frac{u_{A}(x)}{u_{D}(x)}, \\
\label{eq:RAH}
R^{-}_{A/H} &= \frac{\sigma^{DY}(\pi^- + A)}{\sigma^{DY}(\pi^- + \text{H})}
\approx \frac{u_{A}(x)}{u_p(x)}.
\end{align}
%
The up and down nuclear quark distributions are labeled by $u_A$ and 
$d_A$ respectively, $u_{D}$ is the up quark distribution in the deuteron
and $u_p$ the up quark distribution in the proton.
Eqs.~\eqref{eq:Rpm}--\eqref{eq:RAH} demonstrate
that these Drell-Yan cross-section ratios are very sensitive to the
flavor dependence of the EMC effect. Moreover, these ratios are not 
sensitive to the uncertainty of the pion structure functions, which
are not yet determined accurately.

To study the sensitivity of the pion-induced Drell-Yan processes to
the flavor dependence of the EMC effect we calculate these ratios
using the nuclear and nucleon PDFs from the CBT model~\cite{ianemc,ian}. 
Rather than using the approximate expressions of Eqs.~\eqref{eq:Rpm}--\eqref{eq:RAH}, 
the expression for the Drell-Yan cross-section in Eq.~\eqref{eq:crosssection} is
used in the calculations. Terms involving heavy quarks $(s,c,b,t)$ are not 
included due to their negligible contributions. 
The pion-induced Drell-Yan cross-section for each nuclear target is determined
using the nuclear PDFs from the CBT model. 
The free nucleon PDFs obtained from the CBT model are used to calculate the 
Drell-Yan cross-section from the deuteron, with $u_D = (u_p + d_p)/2 = d_D$. 
For the pion PDFs, we use the GRV-P LO parametrization~\cite{pionpdf}. 
As expected, the results are found to be insensitive to the choice of the 
pion PDFs.

\begin{figure}[tbp]
\centering\includegraphics[width=\columnwidth]{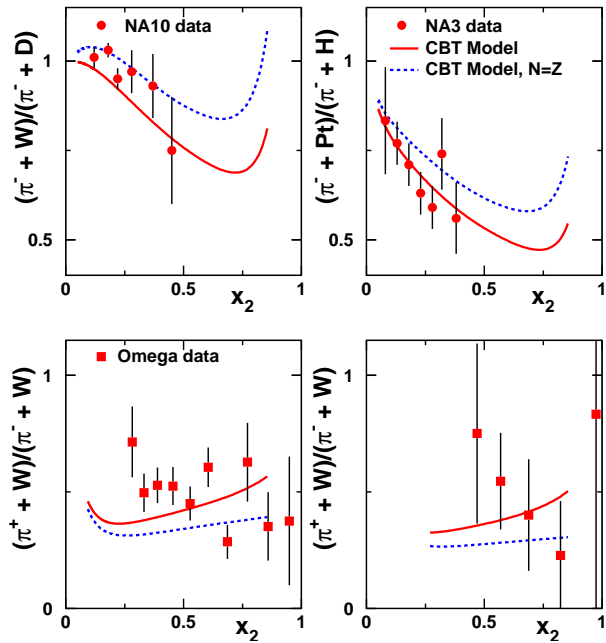}
\caption{The existing data for the ratios $\frac{\sigma^{DY}(\pi^-+\text{W})}
{\sigma^{DY}(\pi^-+\text{D})}$ (upper left), $\frac{\sigma^{DY}(\pi^-+\text{Pt})}
{\sigma^{DY}(\pi^-+\text{H})}$ (upper right) and $\frac{\sigma^{DY}(\pi^++\text{W})}
{\sigma^{DY}(\pi^-+\text{W})}$ (lower) versus the predictions using nuclear PDFs 
of the CBT model for tungsten (red solid) and $N=Z$ nuclear matter 
(blue dashed).}
\label{fig2}
\end{figure}

In Fig.~\ref{fig2} we compare our calculations of the pion-induced
Drell-Yan cross-section ratios with the existing data. The top
left panel shows the ratio of  $\frac{\sigma^{DY}(\pi^-+\text{W})}{\sigma^{DY}(\pi^-+\text{D})}$ 
from the NA10 experiment~\cite{na10}. These plots contain both the
$P_{beam} = 286$ and $140\,$GeV data sets, which are very similar.
Our calculations are performed at
$P_{beam} = 286\,$GeV, since most of the data was obtained at this energy.
We use the PDFs of the CBT model~\cite{ianemc,ian} at a fixed $Q^2$ of $25\,$GeV$^2$, 
which is approximately the mean
$Q^2$ of the NA10 experiment. The top right panel shows
the ratio $\frac{\sigma^{DY}(\pi^-+\text{Pt})}{\sigma^{DY}(\pi^-+\text{H})}$ from the
NA3 experiment~\cite{na3}. The data was collected using a $150\,$GeV $\pi^-$ beam
and the $Q^2$ range covered was $16.8 \leq Q^2\leq70.6\,$GeV$^2$. Our
calculations are performed for $P_{beam} = 150\,$GeV and $Q^2 = 25\,$GeV$^2$. 

The solid curves in Fig. 2 are calculations using the flavor-dependent 
$u_A(x)$ and $d_A(x)$ from the CBT model with $N/Z=1.5$, corresponding
approximately to the $N/Z$ values for the Au, W and Pt nuclei. 
The dashed curves correspond to the calculated ratios using the nuclear 
PDFs from the CBT model for $N=Z$. Since $u_A/u_D = d_A/d_D$ in this case, 
the dashed curves are representative of the predictions for flavor-independent
EMC models. Figure 2 shows that the NA10 data do not exhibit a clear 
preference for the flavor-dependent versus flavor-independent nuclear PDFs.
In contrast, the NA3 data strongly favor the calculations using 
flavor-dependent nuclear PDFs.

\begin{figure}
\centering\includegraphics[width=\columnwidth]{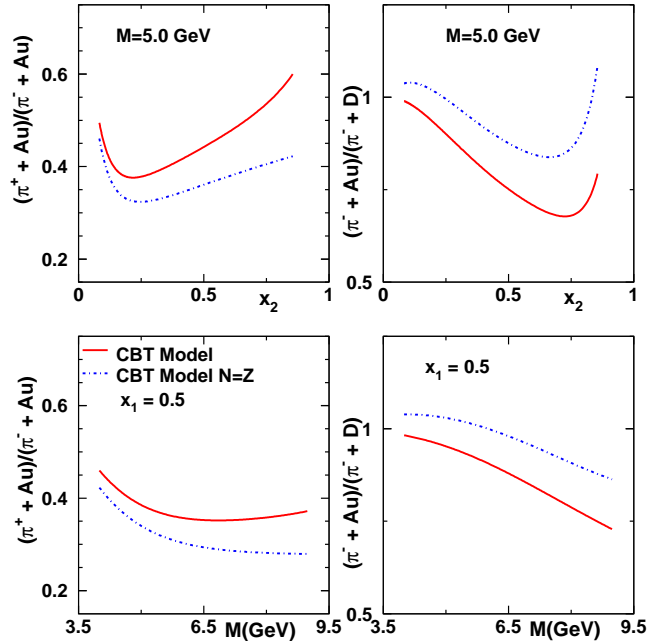}
\caption{The ratio of Drell-Yan cross sections for
$\frac{\sigma^{DY}(\pi^++\text{Au})}{\sigma^{DY}(\pi^-+\text{Au})}$ (upper left)
and $\frac{\sigma^{DY}(\pi^-+\text{Au})}{\sigma^{DY}(\pi^-+\text{D})}$ (upper right)
as a function of $x_2$, using nuclear PDFs from the CBT model
for gold (red solid) and for $N=Z$ nuclear matter (blue dot-dashed).
These calculations are performed for a pion beam energy of 160 GeV,
at $Q^2 = 25\,$GeV$^2$. The lower panels show the same ratios but
as a function of the dimuon mass, $M$, at a fixed $x_1 = 0.5$.}
\label{fig3}
\end{figure}

The Drell-Yan ratio, $R_\pm$, of Eq.~\eqref{eq:Rpm} is an ideal experimental 
observable to search for flavor-dependent EMC effect, since it is directly 
proportional to $d_A(x)/u_A(x)$. The only existing data on $R_\pm$ is from the Omega 
collaboration~\cite{omega}. The lower panels of Fig. 2 show the ratios of 
$\frac{\sigma^{DY}(\pi^++\text{W})}{\sigma^{DY}(\pi^-+\text{W})}$ collected 
with $39.5\,$GeV pions over a $Q^2$ range of $5.3 \leq Q^2 \leq 7.3\,$GeV$^2$
and $16 \leq Q^2 \leq 25\,$GeV$^2$. The calculations were performed at
$Q^2 = 7\,$GeV$^2$ and $Q^2 = 20\,$GeV$^2$, respectively. 
Unfortunately, the existing data lack the precision for placing a useful 
constraint on the flavor dependence of the EMC effect.

The plan for the COMPASS collaboration~\cite{compass} to measure 
Drell-Yan cross sections with $160\,$GeV pion beams offers an exciting
opportunity for a precise test of flavor-dependent EMC effect. Figure 3
shows the predictions of $R_\pm$ and $R_{A/D}^-$ for the COMPASS kinematic
coverage using the PDFs from the CBT model. The upper panels show the
predictions as a function of $x_2$ for a fixed dimuon mass of $5\,$GeV,
while the calculations for a fixed $x_1 = 0.5$ are shown in the lower 
panels as a function of dimuon mass. The significant difference between
the predicted ratios using the flavor-dependent versus flavor-independent
nuclear PDFs provide a strong motivation for such measurements in the future.

It is worth noting that several parametrizations of the nuclear
parton distributions have been obtained from global analyses of
DIS and Drell-Yan data~\cite{eskola1,eskola2,kumano1,kumano2}. However, pion-induced 
Drell-Yan data were not included in these global analyses. The sensitivities
of the pion-induced Drell-Yan data to the flavor dependence of the
nuclear PDFs, as shown in this study, strongly suggest that these
data should be included in future global analyses.

In conclusion, we have examined the possibility of using pion-induced 
Drell-Yan processes as a sensitive experimental tool to study the
flavor dependence of the modification of quark distributions in the nuclear
medium. We suggest several Drell-Yan cross-section ratios 
as the most sensitive experimental quantities for such a study. The existing
Drell-Yan data are not sufficiently accurate yet, although the NA3 data
clearly favor the flavor-dependent over the flavor-independent nuclear PDFs.
Precise future pion-induced Drell-Yan experiments can provide
unique constraints that will help distinguish the various theoretical models 
and most importantly shed new light on the origins of the EMC effect.

This work was supported in part by the U.S. Department of Energy 
and the National Science Foundation.

\hyphenation{Post-Script Sprin-ger}

\end{document}